\documentclass[
               DIV=15, font size=10.5pt,
                 onecolumn,
               ]{scrartcl}  

\usepackage[dvipsnames]{xcolor}
\usepackage{psfrag} 
\usepackage{subcaption}

\usepackage{hyperref}
\usepackage{amsmath}
\usepackage{amssymb}
\usepackage{graphics}
\usepackage{mathtools}
 \usepackage{pstool} 
\usepackage{todonotes}
\usepackage{gensymb}
\usepackage{bm}
\usepackage{multirow}
\usepackage{cancel}
\usepackage{float}
\usepackage{textcomp}
\usepackage{xcolor}
\usepackage[square, numbers,sort&compress]{natbib}
\usepackage[inline]{enumitem}
\usepackage{dirtytalk}
\usepackage[utf8]{inputenc}
 
\usepackage{pgfplots}
\usepackage{pgfplotstable}
\usepgfplotslibrary{groupplots}
\usepackage{filecontents}
\usepackage{algorithm2e}
\usepackage{algorithmic}
\usepackage{siunitx}
\usepackage{tikz}
\usepackage{tikzscale}
\usepackage{tabularx} 
\usetikzlibrary{spy}
\usetikzlibrary{snakes,arrows,shapes}    
\usetikzlibrary{spy}
\usetikzlibrary{backgrounds}  
\usetikzlibrary{decorations}
\usetikzlibrary{positioning}
\usetikzlibrary{patterns}
\usetikzlibrary{calc} 

\usepackage{booktabs}   
\usepackage{makecell} 
\usepackage{colortbl}   

\usepackage{xspace}   
\usepackage{setspace}   
 
\usepackage{soul}
\usepackage{ulem}
\usepackage{currfile}
\usepackage{import}

\usepackage{authoraftertitle} 
\usepackage{authblk} 
\usepackage{cleveref}

\usepackage[english]{babel}

\pgfplotsset{compat=1.8}

\newcommand{\findmax}[3]{
    \pgfplotstablesort[sort key={#2},sort cmp={float >}]{\sorted}{#1}%
    \pgfplotstablegetelem{0}{#2}\of{\sorted}%
    \let #3=\pgfplotsretval%
}

\pgfplotscreateplotcyclelist{myCycleListBlackAndWhite}
{%
  {black, mark=o},
  {black, mark=square},
  {black, mark=triangle},
  {black, mark=diamond}, 
  {black, mark=pentagon}, 
  {black, mark=*},
  {black, mark=square*},
  {black, mark=triangle*},
  {black, mark=diamond*}, 
  {black, mark=pentagon*}, 
}

\definecolor{darkgreen}{rgb}{0,0.4,0} 
\definecolor{darkbrown}{rgb}{0.5, 0.396, 0.09}

\definecolor{c1}{rgb}{0.0, 0.4196078431372549, 0.6431372549019608}
\definecolor{c2}{rgb}{1.0, 0.5019607843137255, 0.054901960784313725}
\definecolor{c3}{rgb}{0.6705882352941176, 0.6705882352941176,
0.6705882352941176} \definecolor{c}{rgb}{0.34901960784313724, 0.34901960784313724, 0.34901960784313724}
\definecolor{c4}{rgb}{0.37254901960784315, 0.6196078431372549,
0.8196078431372549} \definecolor{c}{rgb}{0.7843137254901961, 0.3215686274509804, 0.0}
\definecolor{c5}{rgb}{0.5372549019607843, 0.5372549019607843,
0.5372549019607843} \definecolor{c}{rgb}{0.6352941176470588, 0.7843137254901961, 0.9254901960784314}
\definecolor{c6}{rgb}{1.0, 0.7372549019607844, 0.4745098039215686}
\definecolor{c7}{rgb}{0.8117647058823529, 0.8117647058823529,
0.8117647058823529}

\pgfplotscreateplotcyclelist{myCycleListColor}
{%
  {blue, mark=o},
  {red, mark=square},
  {darkbrown, mark=triangle},
  {darkgreen, mark=diamond},
  {black, mark=pentagon},
  {blue, mark=*},
  {red, mark=square*},
  {darkbrown, mark=triangle*},
  {darkgreen, mark=diamond*},
  {black, mark=pentagon*},
}

\pgfplotsset{every axis/.append style= 
              {
                font=\small,
                mark size=2,
                line width = 0.1,
                legend style={font=\small, mark size=3, draw=none, fill=none},
                legend cell align=left,
                cycle list name=myCycleListColor,
              }
            }
            
\pgfplotstableset
{
  every odd row/.style={before row={\rowcolor[gray]{0.9}}},
  every head row/.style=
  {
    before row=
    {%
      \toprule
    },
    after row=\midrule
  },
  every last row/.style=
  {
    after row=\bottomrule
  }
}

\newif\ifdrawboundingbox

\usetikzlibrary{external} 
\tikzset{external/system call={pdflatex \tikzexternalcheckshellescape
-halt-on-error -interaction=batchmode -jobname "\image" "\texsource"}} 

\pgfkeys
{ 
  /pgf/number format/.cd,
  fixed,
  set thousands separator={\,},
  1000 sep in fractionals,
}

\newcolumntype{C}[1]{>{\centering\arraybackslash}m{#1}}
\newcolumntype{R}[1]{>{\raggedright\arraybackslash}m{#1}}
\newcolumntype{L}[1]{>{\raggedleft\arraybackslash}m{#1}}

\newcommand{\delete}[1]{\xspace}

\definecolor{Reviewer1}{rgb}{0.0, 0.0, 1.0}
\definecolor{Reviewer2}{rgb}{0.0, 0.5, 0.0}

\title{Two-scale analysis and design of spaceframes with complex additive manufactured nodes}

\author[1]{O. Oztoprak\thanks{\href{mailto:oguz.oztoprak@tum.de}{\texttt{oguz.oztoprak@tum.de}},
    Corresponding author}}
\author[2]{A. Paolini}
\author[3]{P. D'Acunto}
\author[1,4]{E. Rank}
\author[1]{S. Kollmannsberger}

 \affil[1]{Chair of Computational Modeling and Simulation,
 Technische Universit\"at M\"unchen, Germany}
 \affil[2]{SOFiSTiK AG, Garching, Germany}
 \affil[3]{Professorship of Structural Design, Technische Universit\"at M\"unchen, Germany}
 \affil[4]{Institute for Advanced Study, Technische Universit\"at M\"unchen, Germany}

\newcommand{\journal}{Engineering Structures}
\newcommand{\publicationDate}{\today}

\usepackage{fancyhdr}
\date{}
\fancypagestyle{plain}{%
  \fancyhf{}%
  \fancyfoot[L]{
  \vspace{-1.25cm} 
  \footnotesize{Preprint submitted to \textit{\journal{}}  \hfill
  \publicationDate
  \\
  }} }

\newtheorem{theorem}{Theorem}[section]
\newtheorem{remark}[theorem]{Remark}

\usepackage[dvipsnames]{xcolor}
\colorlet{Reviewer1}{Green}
\colorlet{Reviewer2}{Blue}
\colorlet{Reviewer3}{Red}
\colorlet{Reviewer4}{Fuchsia}
\usepackage{ulem}
\makeatletter
\@addtoreset{footnote}{section}
\makeatother
\crefname{figure}{Fig.}{Fig.}
\crefname{equation}{Eq.}{Eq.}
\crefname{table}{Tab.}{Tab.}

\drawboundingboxtrue
\drawboundingboxfalse 
\newcommand*{\figref}[2][]{%
	\hyperref[{fig:#2}]{%
		Fig.~\ref*{fig:#2}%
		\ifx\\#1\\%
		\else
		\,#1%
		\fi
	}%
}
\definecolor{changes}{RGB}{0,0,0}
\definecolor{changez}{RGB}{0,0,0}

\begin{document}  

\normalem
\maketitle  
  
\vspace{-1.5cm} 
\hrule 
\section*{Abstract}
{The advancements in additive manufacturing (AM) technology have allowed for the production of geometrically complex parts with customizable designs. This versatility benefits large-scale space-frame structures, as the individual design of each structural node can be tailored to meet specific mechanical and other functional requirements. To this end, however, the design and analysis of such space-frames with distinct structural nodes needs to be highly automated. A critical aspect in this context is automated integration of the local 3D features into the 1D large-scale models. In the present work, a two-scale modeling approach is developed to improve the design and linear-elastic analysis of space frames with complex additively manufactured nodes. The mechanical characteristics of the 3D nodes are numerically reduced through an automated dimensional reduction process based on the Finite Cell Method (FCM) and substructuring. The reduced stiffness quantities are assembled in the large-scale 1D model which, in turn, enables efficient structural analysis. The response of the 1D model is passed on to the local model, enabling fully resolved 3D linear-elastic analysis. The proposed approach is numerically verified on a simplified beam example. Furthermore, the workflow is demonstrated on a tree canopy structure with additively manufactured nodes with bolted connections. The form of the large-scale structure is found based on the Combinatorial Equilibrium Modeling framework, and the different designs of the local structural nodes are based on generative exploration of the design space. It is demonstrated that the proposed methodology effectively automates the design and analysis of space-frame structures with complex, distinct structural nodes.}

\vspace{.2cm} 
\vspace{0.25cm}
\noindent \textit{Keywords:} Additive Manufacturing, Multiscale problems, Finite Cell Method, Frame structures,  Finite Element Method
 
\vspace{0.35cm}
\hrule 
\vspace{0.15cm}
\captionsetup[figure]{labelfont={bf},name={Fig.},labelsep=colon}
\captionsetup[table]{labelfont={bf},name={Tab.},labelsep=colon}
	\section{Introduction} \label{sec:intro}
Additive manufacturing (AM) offers enhanced geometric freedom in the design of highly customized structural parts with favorable features. Structural nodes, for example, have been additively manufactured to achieve the desired performance while remaining lightweight \cite{galjaard2015}. Such nodes can be used in tensegrity \cite{sultan2003} and space frame structures \cite{lan1999}, which are found in a wide range of industrial applications. Contrary to the parts manufactured by conventional techniques, AM components have the additional advantage that their highly customizable design can satisfy varying mechanical requirements in different areas of construction.

The practical implication of the various advantages and recent developments in additive manufacturing is the growing use of these parts in large-scale constructions \cite{pacillo2021}. Yet, the conventional analysis and design approaches are not well suited to incorporate such complex structural parts with unique and individual designs into a seamless and efficient workflow \cite{paolini2019}. This paper presents a method for a tighter integration between conventional structural analysis and individualized AM parts in the design and analysis of engineering structures.

A common practice in structural engineering is to simplify geometric details in the global analysis of large-scale structures, leading to an efficient computation with reasonable accuracy. In other words, considering a finite element (FE) based structural analysis, it is impractical, if not impossible, to spatially resolve the entire structure with 3D elements in full detail. Multiscale modeling techniques address this challenge by partitioning the large-scale structure into interrelated, smaller-scale components which are modeled depending on the component's physical complexity.

There exists a considerable amount of research on the multiscale modeling of structures, which can broadly be divided into two groups following \cite{fish2011}: \begin{enumerate*}[label=(\alph*)]
    \item simultaneous and 
    \item information-transferring 
\end{enumerate*} modeling. Simultaneous multiscale approaches resolve the existing geometric scales in one unified model. A subgroup of simultaneous multiscale approaches is mixed dimensional modeling where the coupling of mixed-dimensional finite elements such as beam-to-shell or beam-to-solid occurs in a single model. The coupling on the element interfaces ensures displacement compatibility and stress equilibrium. To this end, multi-point constraints \cite{mccune2000}, Lagrange multipliers \cite{halliday1999, romero2018}, Nitsche's method \cite{yamamoto2019} or the Arlequin method \cite{dhia2005} can be employed. Methods that utilize the superposition of mathematical models \cite{Krause2003, Duster2007} can also be classified within this category, where the solution space $u$ is decomposed into local and global contributions $u_{\text{local}}$ and $u_{\text{global}}$, resulting in $u = u_{\text{global}} + u_{\text{local}}$. In this case, prescribing homogeneous boundary conditions on $u_{\text{local}}$ at the local-global interface ensures the solution compatibility. On the other hand, information-transferring approaches often have a separate model for the fine geometric scale, whose response is computed under certain assumptions and transferred to the large-scale model. Numerical homogenization techniques based on the representative volume elements (RVEs) are such approaches that perform numerical material characterization analysis and infuse the material response into the large-scale model \cite{korshunova2020,korshunova2021a,sciegaj2020}.

The present work proposes a two-scale, information-transferring approach for the design and linear-elastic analysis of space frame structures with additively manufactured parts, where special emphasis is placed on the aspect of automatization. The global scale model assumes 1D linear beam elements, whereas on the local-scale structural nodes are modeled based on fully resolved 3D models. The relevant mechanical characteristics of complex structural nodes are numerically reduced independent of the global scale configuration, following an automated dimensional reduction process, which can be interpreted as substructuring. The reduced stiffness quantities are transferred in the local-to-global direction, and efficient structural analysis of the global scale is facilitated. The global solution at the interface of the scales is then imposed on the local-scale models, where the complete 3D stress state of the structural nodes can be analyzed. The Finite Cell Method (FCM) is employed for the analysis on the local-scale due to its ability to easily and seamlessly incorporate complex geometries. The iterative engineering design and analysis workflow greatly benefits from the FCM, as it can reduce the amount of manual engineering labor associated with the two-scale approach. The advantages of the proposed approach are illustrated on a tree canopy structure with additively manufactured structural nodes. The global design of the structure is provided as a result of a form-finding based on the Combinatorial Equilibrium Modeling framework \cite{ohlbrock2020}. By contrast, the local structural nodes are generatively designed to explore the design space for alternative feasible designs swiftly.

The structure of this contribution is as follows. \Cref{sec:methods} presents the core concepts of the proposed two-scale modeling approach, including the overview in~\Cref{sec:two-scale}, the details of the global scale modeling in~\Cref{sec:global-scale}, and the modeling techniques used at the local-scale in~\Cref{sec:local-scale}. Numerical examples are presented in~\Cref{sec:cantilever}, with a simple numerical example used to verify the methodology in~\Cref{sec:cantilever} and a real-life application demonstrating the complete workflow in~\Cref{sec:complextree}.

\begin{figure}
	\hspace{1cm}
	\centering
	\includegraphics[width=1.0\textwidth]{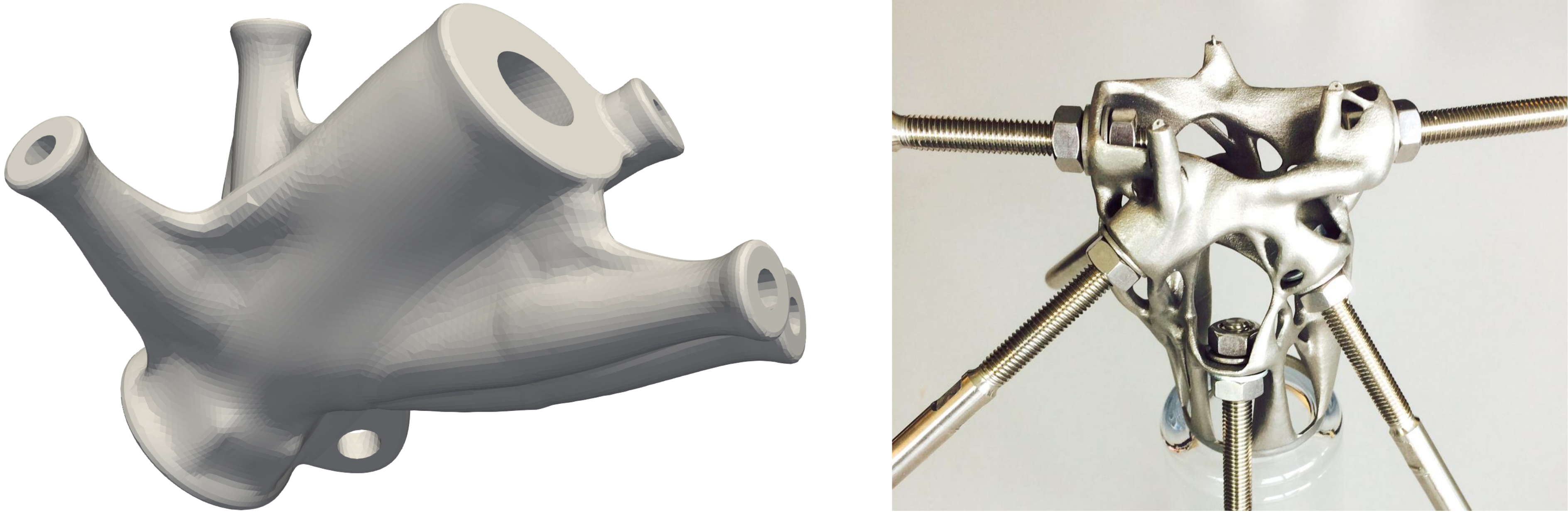}
    \caption{{Additively manufactured nodes illustrating the potential geometric freedom: left:\cite{ghantasala2021}, right:\cite{galjaard2015}}}
	\label{fig:am+nodes}
\end{figure}

	\section{Methods} \label{sec:methods}

\subsection{Two-scale model} \label{sec:two-scale}
The terms \textit{microscale} and \textit{macroscale} are commonly used in multiscale modeling literature to distinguish the two associated geometric scales. In this contribution, the \textit{microscale} is referred to as \textit{local-} or \textit{part-scale}, whereas the \textit{macroscale} is indicated as \textit{global-} or \textit{large-scale}.

The geometric details of the part-scale are often not fully resolved during a computationally efficient structural analysis. As a remedy, the whole structure can be partitioned into components depending on the existing level of detail (e.g. geometric detail) required to model the component, as in~\cite{mata2008}. In the case of a space frame, such a partitioning leads to two distinct sets of structural elements:
\begin{enumerate}[label=(\arabic*)]
    \item The set of beams $\beta_{m}$ that can readily be represented by 1D models.
    \item The set of structural nodes $\Omega_{ns}$ which connect the beams. They typically have complex geometries and their mechanical properties are, in general, difficult to incorporate in a large-scale model.
\end{enumerate}
        In general, each structural node $\omega_{n} \in \Omega_{ns}$ has a distinct material distribution and internal structure, resulting in complex models whose mechanical analysis requires the evaluation of a fully resolved 3D problem. 

\begin{figure}[h!]
	\centering
	\includegraphics[width=0.65\textwidth]{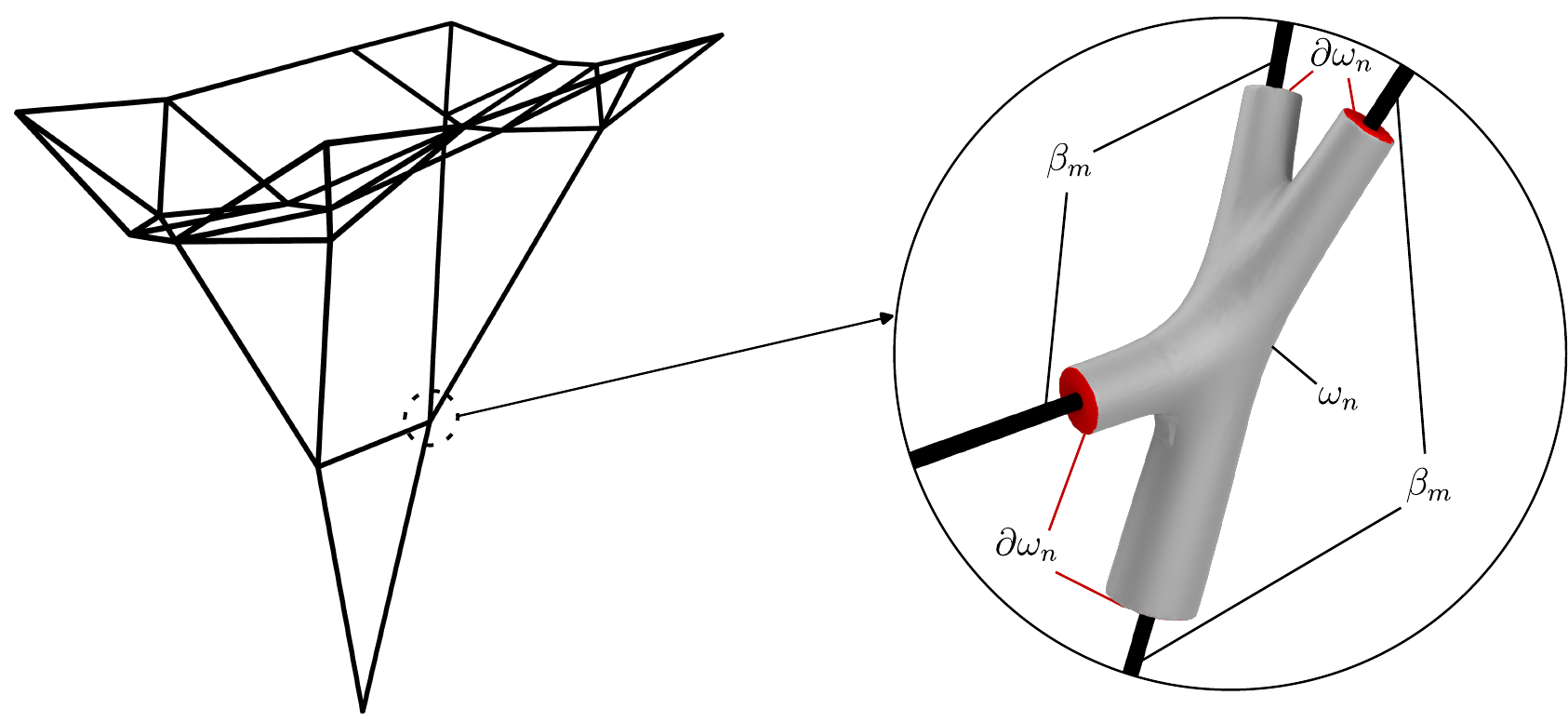}
    \caption{{The illustration of the two-scale model: the space-frame in the global-scale~\cite{pastrana2022} (left) and the structural node at the part-scale with its interface to the beam elements (right).}}
	\label{fig:two+scale}
\end{figure}

        The interface surfaces between $\beta_{m}$ and a particular node $\omega_{n}$ with $p$ connections is defined as $\partial\omega_{n} = \bigcup_{i=1}^{p}\partial\omega_{ni}$. Each surface in $\partial\omega_{n}$ is an interface to a beam element in $\beta_{m}$, as shown in~\cref{fig:two+scale} where the beam elements are represented in black and the beam-node interfaces are shown in red. The beam degrees of freedom on these interfaces will be referred to as the boundary degrees of freedom. The coupling of the scales is done by dimensionally reducing the 3D part-scale models of $\omega_{n}$ to 0D models in such a way that the reduced models can properly be assembled into the large-scale model. This reduction must ensure that the kinematic assumptions of the reduced 0D models are consistent with the assumptions of the large-scale model. In this contribution, the Timoshenko beam finite elements~\cite{cowper1966} are used in the large-scale model. Therefore, the reduced models of $\Omega_{ns}$ are determined with respect to the degrees of freedom of the Timoshenko beam elements. The key assumptions are:
\begin{enumerate*}[label=(\alph*)]
    \item The deformations are small in both scales.
    \item The plane sections in the large-scale model and on the interfaces $\partial\omega_{n}$ remain plane after deformation.
\end{enumerate*}

The dimensional reduction of each $\omega_{n}$ model closely follows the static condensation procedure~\cite{paz2001}, where a change of basis results in a reduced-dimensional representation of the original stiffness matrix. Details regarding the computation of the change-of-basis matrix (transformation matrix) are described in~\Cref{sec:local-scale}. The reduced stiffness matrix is then handled by the large-scale model as a ‘superelement’ during the assembly. In other words, the large-scale model considers each $\omega_{n}$ as a 0D element that stems from a dimensional reduction process. This process can be interpreted as a projection of the degrees of freedom of the 3D node model onto a 0D element in the global-scale. The entries of the 0D element do not add additional degrees of freedom to the global system matrix. Instead, they can be interpreted as a type of nodal stiffness which defines the magnitude of the displacements of $\partial\omega_{n}$ per unit force relative to each other. Note that the static condensation process of each node is independent of the rest of the structure, i.e. the reduced model depends entirely on the physical properties of the structural node.   

The first part of the two-scale model follows the local-to-global direction using the dimensional reduction process introduced above. The second part propagates the solution from the large-scale model to the part-scale model to perform a stress state analysis on the fully resolved local model. To this end, the displacements at the boundary degrees of freedom are imposed on the fully resolved local model to solve the local, linear elastic problem.

\subsection{Global Scale Model}  \label{sec:global-scale}
\subsubsection{The beam model}  \label{sec:beam-model}
\begin{figure}[h!]
	\centering
	\includegraphics[width=0.7\textwidth]{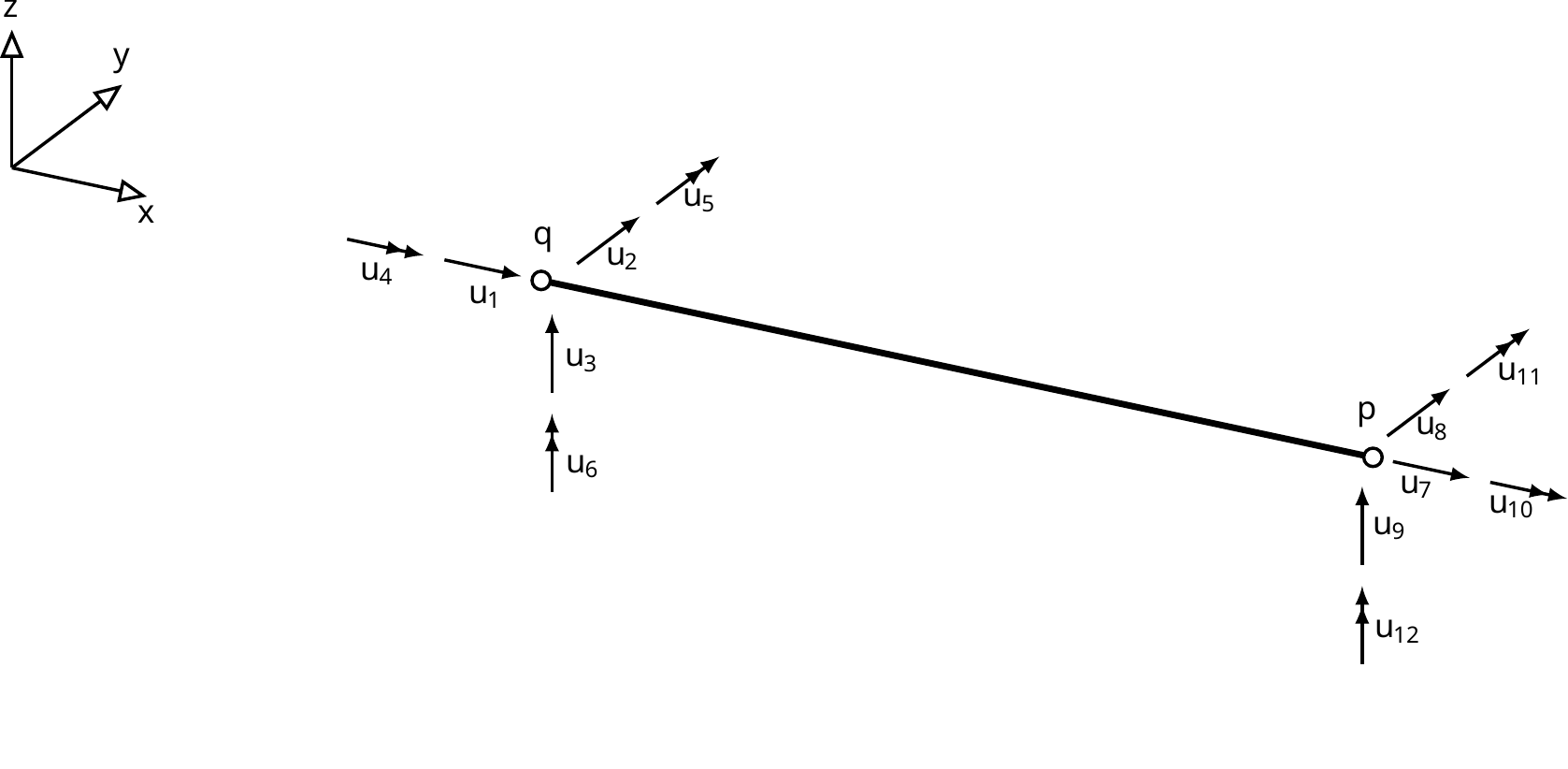}
	\caption{{The beam element}}
	\label{fig:timo+beam}
\end{figure}

The global-scale structure is modeled by 2-node Timoshenko beam finite elements in 3D with 6 degrees of freedom (DOF) per node, these include the axial displacements ($u_{1}$, $u_{7}$), the shear displacements ($u_{2}$, $u_{3}$, $u_{8}$, $u_{9}$), the bending rotations ($u_{5}$, $u_{6}$, $u_{11}$, $u_{12}$), and the twisting rotations ($u_{4}$, $u_{10}$), as depicted in~\cref{fig:timo+beam}. The element equilibrium equation is provided below, following~\cite{przemieniecki1968}.
\begin{equation}
    \boldsymbol{K_{e}}\boldsymbol{u_{e}} = \boldsymbol{f_{e}}
\end{equation}
where the element stiffness matrix reads:

\begin{equation}\label{eq:timo+beam}
\setlength{\arraycolsep}{1.8pt}
    \boldsymbol{K_{e}} = \begin{bmatrix}
    \frac{EA}{L} &  &  &  &  &  &  &  &  &  &  &  \\
    0 & \frac{12EI_{z}}{\hat\phi_{z}L^{3}} &  &  &  &  &  &  &  &  &  & \\
    0 & 0 & \frac{12EI_{y}}{\hat\phi_{y}L^{3}} &  &  &  &  &  &  &  &  & \\
    0 & 0 & 0  & \frac{GJ}{L}  &  &  &  &  &  &  &  & \\ 
    0 & 0 &  \frac{-6EI_{y}}{\hat\phi_{y}L^{2}} & 0 & \frac{(4+\phi_{y})EI_{y}}{\hat\phi_{y}L}  &  &  &  &  &  &  & \\
    0 & \frac{6EI_{z}}{\hat\phi_{z}L^{2}} & 0 & 0 & 0  &\frac{(4+\phi_{z})EI_{z}}{\hat\phi_{z}L}  &  &  &  &  &  & \\
    \frac{-EA}{L} & 0  & 0  & 0  & 0 & 0 & \frac{EA}{L}  &  &  &  & &  \\
    0 & \frac{-12EI_{z}}{\hat\phi_{z}L^{3}}  & 0  & 0  & 0 & \frac{-6EI_{z}}{\hat\phi_{z}L^{2}} & 0 & \frac{12EI_{z}}{\hat\phi_{z}L^{3}}  &  &  & &  \\
    0 & 0  & \frac{-12EI_{y}}{\hat\phi_{y}L^{3}}  & 0  & \frac{6EI_{y}}{\hat\phi_{y}L^{2}} & 0 & 0 & 0  & \frac{12EI_{y}}{\hat\phi_{y}L^{3}} &  & &  \\
    0 & 0  & 0  & \frac{-GJ}{L}  & 0 & 0 & 0 & 0 & 0 & \frac{GJ}{L} & &  \\
    0 & 0  & \frac{-6EI_{y}}{\hat\phi_{y}L^{2}}  & 0  & \frac{(2-\phi_{y})EI_{y}}{\hat\phi_{y}L} & 0 & 0 & 0 & \frac{6EI_{y}}{\hat\phi_{y}L^{2}} & 0 & \frac{(4+\phi_{y})EI_{y}}{\hat\phi_{y}L} &  \\
    0 & \frac{6EI_{z}}{\hat\phi_{z}L^{2}}  & 0  & 0  & 0 & \frac{(2-\phi_{z})EI_{z}}{\hat\phi_{z}L} & 0 & \frac{-6EI_{z}}{\hat\phi_{z}L^{2}} & 0 & 0 & 0 & \frac{(4+\phi_{z})EI_{z}}{\hat\phi_{z}L}  
\end{bmatrix}
\end{equation}
with 
\begin{equation}
    \phi_{z} = \frac{12EI_{z}}{\kappa GAL^{2}}
    \hspace{1cm}
    \hat\phi_{z} = 1 + \phi_{z}
    \hspace{1cm}
    \phi_{y} = \frac{12EI_{y}}{\kappa GAL^{2}}
    \hspace{1cm}
    \hat\phi_{y} = 1 + \phi_{y}.
\end{equation}
The nodal displacement vector $\boldsymbol{u_{e}}$ and the corresponding nodal force vector $\boldsymbol{f_{e}}$ are defined as:
\begin{align}
    \boldsymbol{u_{e}} = \begin{bmatrix}
        u_{1} \\
        u_{2} \\
        u_{3} \\
        u_{4} \\
        u_{5} \\
        u_{6} \\
        u_{7} \\
        u_{8} \\
        u_{9} \\
        u_{10} \\
        u_{11} \\
        u_{12} 
    \end{bmatrix}
    \hspace{3cm}
    \boldsymbol{f_{e}} = \begin{bmatrix}
        f_{1}  \\
        f_{2}  \\
        f_{3}  \\
        f_{4}  \\
        f_{5}  \\
        f_{6}  \\
        f_{7}  \\
        f_{8}  \\
        f_{9}  \\
        f_{10} \\
        f_{11} \\
        f_{12} 
    \end{bmatrix}
\end{align}

The shear correction factor $\kappa$ is calculated for different cross-sections following~\cite{cowper1966}. For a more detailed derivation, the reader is referred to~\cite{przemieniecki1968,andersen,chinwubaike2019}. The assembly of element stiffness matrices constitute the global stiffness matrix $\boldsymbol{K_{\text{beam-global}}}$. The part-scale model is subsequently introduced.

\subsection{Local-scale model} \label{sec:local-scale}
The structural nodes are analyzed in the part-scale, where the complete CAD geometry of each individual node is considered to model the mechanical problem based on 3D linear elastostatics. The purpose of the local-scale model is two-fold: to dimensionally reduce the geometrically complex node model such that it can be integrated into an efficient structural analysis on the large-scale, and to analyze the final stress state of the nodes considering the loading conditions from the large-scale analysis. For this purpose, we employ the Finite Cell Method (FCM) to automate the analysis in the local-scale. The suitability of the FCM for similar types of linear elastic problems involving additively manufactured parts (e.g.~\cite{korshunova2021,korshunova2021a}) and for various other applications have been extensively investigated for a variety of geometric representations, e.g.~\cite{Duster2008,kudela2020,wassermann2020,Wassermann2019}.
\subsubsection{Overview of the Finite Cell Method}
The basics of the FCM within the context of linear elastic problems are summarized below. The reader is referred to~\cite{Duster2008,duster2017} for further details.

\begin{figure}[h!]
	\centering
    \includegraphics[width=1.00\textwidth]{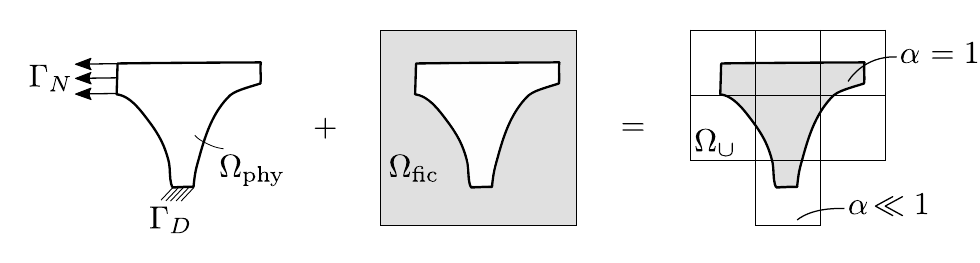}
	\caption{Illustration of the Finite Cell Method~\cite{Parvizian2007}. The cells that do not intersect the physical domain are not considered.}
	\label{fig:fcm+illustration}
\end{figure}

The FCM is an extension to the classical finite elements approach that combines high-order finite elements with an immersed boundary method. The core principle of the FCM is depicted in~\cref{fig:fcm+illustration}. The physical domain $\Omega_{phy}$ is extended by a fictitious domain $\Omega_{fic}$ such that their union $\Omega_{\cup}$ is a trivial shape that can easily be meshed. Thus, the challenge to generate a suitable mesh, which in some cases may require extensive manual labor, is replaced with the resolution of the original domain during numerical integration, where integration schemes such as spacetree subdivision technique can be used for accurate integration. To this end, the recovery of the original problem within $\Omega_{\cup}$ is facilitated by an indicator function $\alpha$ defined as:

\begin{equation}
\label{eq:alpha}
\alpha(\boldsymbol{x})=\begin{cases}
1 & \forall\boldsymbol{x} \in \Omega_{phy} \\
10^{-k} & \forall\boldsymbol{x} \notin \Omega_{phy}.
\end{cases}
\end{equation}
The fictitious domain can physically be interpreted as a very soft material surrounding or filling the voids within the physical domain. It has been shown that for an adequately large $k$ (typically $k$ takes a value within the range of 5 to 10), the existence of the $\Omega_{fic}$ does not add a significant contribution to the error~\cite{Dauge2015a}.
The linear elastic problem can be formulated on $\Omega_{\cup}$ using the following weak form~\cite{Zander2017a}:
\begin{align} \label{eq:linear+elastic+problem}
a(\boldsymbol{u}, \boldsymbol{v}) = f(\boldsymbol{v}). 
\end{align}
The bilinear form reads:
\begin{align} 
    a(\boldsymbol{u}, \boldsymbol{v}) &= \int_{\Omega_{phy}} 1 \  \varepsilon(\boldsymbol{v}):\mathbb{C}:\varepsilon(\boldsymbol{u}) \, \text{d}\Omega + \int_{\Omega_{fic}} \alpha \  \varepsilon(\boldsymbol{v}):\mathbb{C}:\varepsilon(\boldsymbol{u}) \, \text{d}\Omega \label{eq:bilinear-form}\\
    &= \int_{\Omega_{\cup}} \varepsilon(\boldsymbol{v}):\alpha \mathbb{C}:\varepsilon(\boldsymbol{u}) \, \text{d}\Omega , \nonumber
\end{align}
where $\varepsilon$, $\mathbb{C}$, $\boldsymbol{u}$ and $\boldsymbol{v}$ denote the engineering strain tensor, the elastic material tensor, the displacement vector and the vector of test functions, respectively. The right hand side is:
\begin{align} \label{eq:right+hand+side}
    f(\boldsymbol{v}) &= \int_{\Omega_{\cup}} \alpha \,  \boldsymbol{v} \,  \boldsymbol{f} \, \text{d}\Omega + \int_{\Gamma_{N}} \boldsymbol{v} \,  \boldsymbol{t} \, \text{d}\Gamma,
\end{align}
where $\boldsymbol{f}$ denotes the body forces. The traction vector $\boldsymbol{t}$ acts as the Neumann boundary condition on $\Gamma_{N}$. Prescribed displacements $\boldsymbol{u_{p}}$ act on the Dirichlet boundary $\Gamma_{D}$ as:
\begin{align*}
    \boldsymbol{u} = \boldsymbol{u_{p}} \qquad \forall\boldsymbol{x} \in \Gamma_{D}.
\end{align*}
The displacement vector $\boldsymbol{u}$ and the test function $\boldsymbol{v}$ can be approximated in discrete form as a linear combination of the shape functions $N_i$, which are of high order in the case of the FCM, times their corresponding unknown coefficients:
\begin{align} \label{eq:shape+functions}
    \boldsymbol{u} &= \sum_{i}N_i \hat u_i = \boldsymbol{N}\boldsymbol{\hat u} \\
    \boldsymbol{v} &= \sum_{i}N_i \hat v_i = \boldsymbol{N}\boldsymbol{\hat v}
\end{align}
The discrete representation of the problem in~\cref{eq:linear+elastic+problem} is obtained following the Bubnov-Galerkin method~\cite{zienkiewicz2000}: 

\begin{align} \label{eq:k+u+f}
    \boldsymbol{K} \boldsymbol{\hat u} = \boldsymbol{f} 
\end{align}
where $\boldsymbol{K}$, $\boldsymbol{\hat u}$ and $\boldsymbol{f}$ describe the stiffness matrix, the vector of (unknown) displacement coefficients and the load vector, respectively. As the finite cell mesh does not conform to the boundary of the physical domain, the imposition of boundary conditions requires additional care. Homogeneous Neumann boundary conditions demand no special treatment, similar to the classical FEM, while non-homogeneous Neumann boundary conditions can be imposed by computing the second term (boundary integral term) in~\cref{eq:right+hand+side} with the help of an explicit surface discretization of $\Gamma_{N}$. On the other hand, Dirichlet boundary conditions can be imposed in a weak sense by extending the weak form in~\cref{eq:bilinear-form} and~\cref{eq:right+hand+side} with additional constraints. To this end, Nitsche's method~\cite{Nitsche1971} or Penalty method~\cite{Babuska1973} can be employed, e.g.~\cite{Ruess2012}.

\subsubsection{Substructure analysis}\label{sec:substructure-analysis}
The substructuring of the structural nodes is based on the static condensation approach. We formulate the procedure in terms of a reduction of basis. The reduction operation acts on a full dimensional matrix denoted as $\boldsymbol{K}_{n \times n} \in \mathbb{R}^{n \times n}$, which, in our case, represents the unconstrained stiffness matrix of a fully resolved node model in the part-scale. The goal is to determine a reduced dimensional interpretation of $\boldsymbol{K}_{n \times n}$ in terms of the corresponding degrees of freedom in the large-scale model. We refer to this reduced dimensional matrix as the substructure stiffness matrix $\boldsymbol{K}_{k \times k}$ where $k \ll n$.

\begin{figure}[h!]
	\centering
    \includegraphics[width=0.85\textwidth]{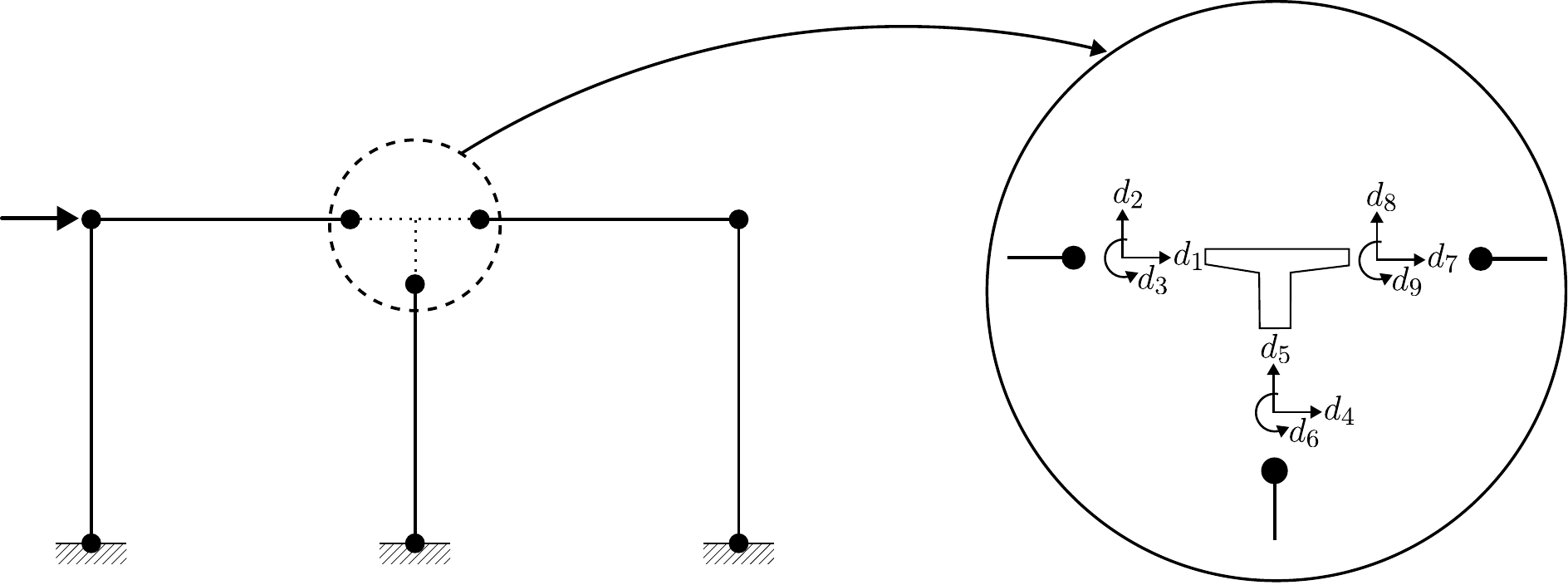}
	\caption{An illustration of the substructuring approach in 2D, left-hand side: the global scale beam model with the reduced structural node, right-hand side: the node in local scale with the boundary degrees of freedom. }
	\label{fig:two+scale+dofs}
\end{figure}

Let $V$ denote a $k$ dimensional vector subspace of $\mathbb{R}^{n}$ with a basis $B = \{\boldsymbol{v_{1}},\, \boldsymbol{v_{2}}, \, \boldsymbol{v_{3}} \quad ... \quad \boldsymbol{v_{k}}\}$ consisting of $k$ basis vectors. Additionally, we define $\boldsymbol{u} \in \mathbb{R}^{n}$ as a displacement vector. We can represent $\boldsymbol{u}$ as:

\begin{equation} \label{eq:u+and+ub}
    \boldsymbol{u} = \boldsymbol{N}_{n \times k} \,  [\boldsymbol{u}]_{B},
\end{equation}
where $[\boldsymbol{u}]_{B}$ is the $k$ dimensional displacement vector with respect to the reduced basis $B$, and $\boldsymbol{N}_{n \times k}$ is a linear map whose columns are the basis vectors $\{\boldsymbol{v_{1}},\, \boldsymbol{v_{2}}, \, \boldsymbol{v_{3}} \quad ... \quad \boldsymbol{v_{k}}\}$ . In our case, $k$ represents the total number of beam degrees of freedom on the beam-node interfaces $\partial\omega_{n}$, as illustrated in~\cref{fig:two+scale+dofs} for a 2D case and in~\cref{fig:two+scale} for a 3D case. As an example, the node in~\cref{fig:two+scale+dofs} interfaces with 3 beams with a total of $k=9$ boundary degrees of freedom. More specifically, the substructure is coupled to the rest of the large-scale structure through the $k$ boundary degrees of freedom and therefore, the $k$ linearly independent basis vectors must be capable of representing the corresponding degrees of freedom. Thus, a column vector $\boldsymbol{v_{i}} \in \boldsymbol{N}_{n \times k}$ has the following physical interpretation: it represents the displacement field of the fully resolved node under a unit deformation in the direction of the corresponding global-scale degree of freedom $i$, while all other degrees of freedom are constrained. Note that the $k$ boundary degrees of freedom in question are defined in the global coordinate system as the substructure has no direct interpretation of a local coordinate system.

\begin{figure}[h!]
	\centering
    \includegraphics[width=0.90\textwidth]{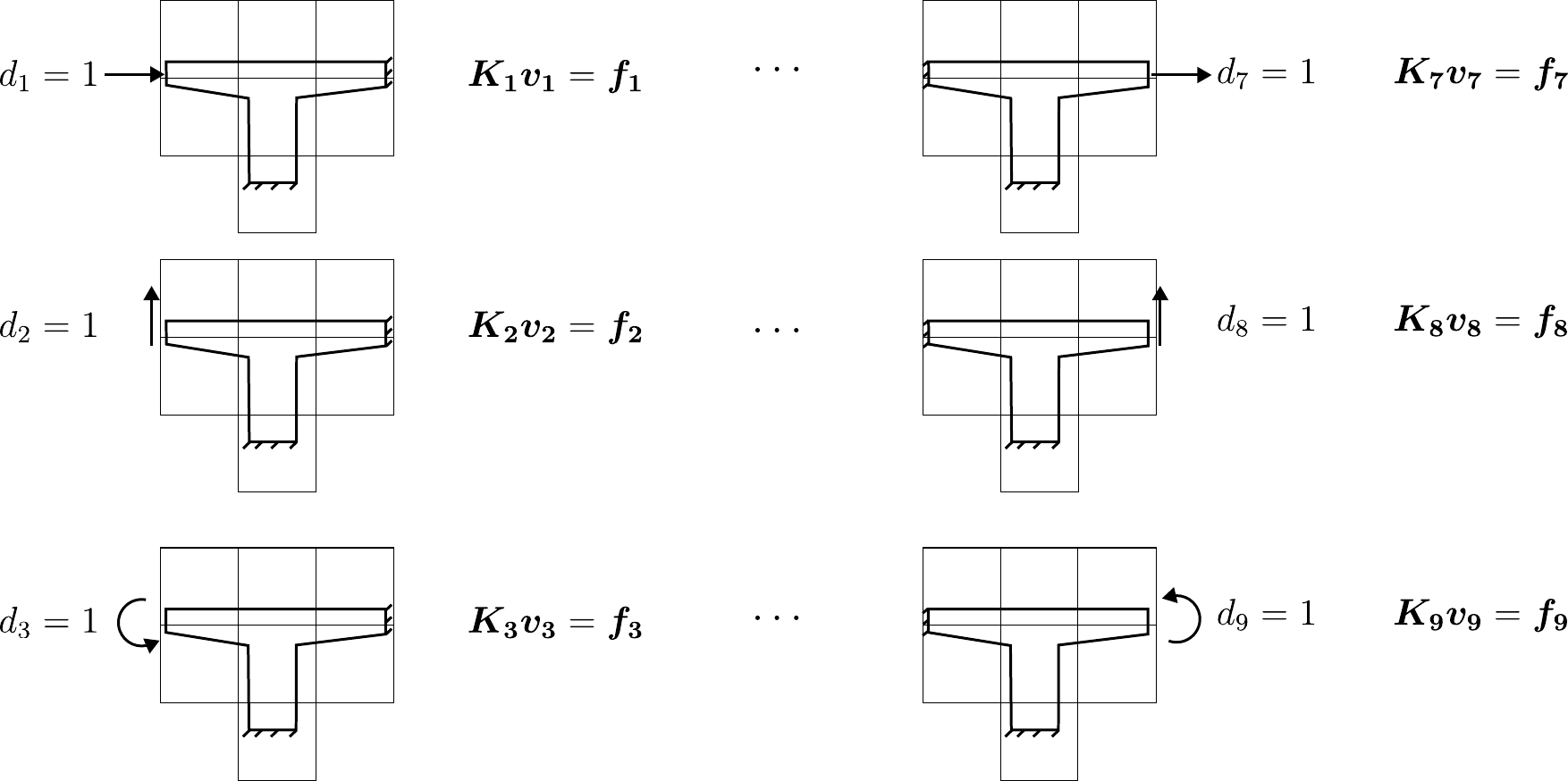}
    \caption{Computation of the change of basis matrix $\boldsymbol{N}_{n \times k}$ for a 2D node from~\cref{fig:two+scale+dofs}. The solutions $\boldsymbol{v_{1}},\, \boldsymbol{v_{2}}, \, \boldsymbol{v_{3}} \quad ... \quad \boldsymbol{v_{9}}$ of the illustrated finite cell problems constitute the columns of the change of basis matrix.}
	\label{fig:two+scale+reduce+basis}
\end{figure}

The substructure stiffness matrix $\boldsymbol{K}_{k \times k}$ can be obtained based on the principle of virtual work~\cite{przemieniecki1968}. Accordingly, the virtual strain energy can be written as:
\begin{equation} \label{eq:virtual+strain+energy}
    \delta{U_{i}} \, = \, \delta{\boldsymbol{u}^{T}}\, \boldsymbol{K}_{n \times n}\, \boldsymbol{u}.
\end{equation}
Plugging~\cref{eq:u+and+ub} into~\cref{eq:virtual+strain+energy} results in:

\begin{align*} \label{eq:virtual+strain+energy}
    \delta{U_{i}} \, &= \, \delta{[\boldsymbol{u}]_{B}^{T}} \ \boldsymbol{N}_{n \times k}^{T} \ \boldsymbol{K}_{n \times n} \ \boldsymbol{N}_{n \times k} \ [\boldsymbol{u}]_{B} \\
     &= \, \delta{[\boldsymbol{u}]_{B}^{T}} \ \boldsymbol{K}_{k \times k} \ [\boldsymbol{u}]_{B}, 
\end{align*}
where the substructure stiffness matrix is defined as:
\begin{equation} \label{eq:substructure+stiffness}
\boldsymbol{K}_{k \times k} = \boldsymbol{N}_{n \times k}^{T} \  \boldsymbol{K}_{n \times n} \  \boldsymbol{N}_{n \times k}.
\end{equation}
\cref{eq:substructure+stiffness} has the canonical structure of a change of basis on a bilinear form~\cite{guyan1965,turner1956,kreutz2013}. Here, the challenge is the computation of the change of basis matrix $\boldsymbol{N}_{n \times k}$, which involves solving $k$ linear elastic problems on the fully resolved model in the part-scale, following the above interpretation of $\boldsymbol{N}_{n \times k}$. The FCM is employed for the aforementioned local-scale computations due to its robust handling of complex geometries, which facilitates the automated dimensional reduction of each individual structural node with different geometrical features. \cref{fig:two+scale+reduce+basis} depicts the linear systems involved in the computation of the nine column vectors $\boldsymbol{v_{1}},\, \boldsymbol{v_{2}}, \, \boldsymbol{v_{3}} \quad ... \quad \boldsymbol{v_{9}}$ of $\boldsymbol{N}_{n \times k}$ for a 2D structural node using FCM, based on the example introduced in~\cref{fig:two+scale+dofs}.

\begin{remark}
    The computation of the unconstrained stiffness matrix $\boldsymbol{K}_{n \times n}$ of a fully resolved 3D model is one of the compute-intensive parts of the workflow. On the other hand, the $k$ linear elastic problems associated with the computation of $\boldsymbol{N}_{n \times k}$ differ only with respect to the boundary conditions as illustrated in~\cref{fig:two+scale+reduce+basis}, i.e. $\boldsymbol{K}_{n \times n}$ remains unchanged. Thus, the unconstrained stiffness matrix associated with a fully resolved part-scale model is computed once and reused across the series of computations. 
\end{remark}

\begin{remark}
     In this contribution, we impose Dirichlet boundary conditions in a weak sense following the Penalty method~\cite{Babuska1973}. This is the case for the local-scale problems associated with the substructure analysis in~\Cref{sec:substructure-analysis} as well as the computation of the final stress state in~\Cref{sec:global-to-local}.
\end{remark}

\subsubsection{Global-to-local transition}\label{sec:global-to-local}
\begin{figure}[h!]
	\centering
    \includegraphics[width=0.93\textwidth]{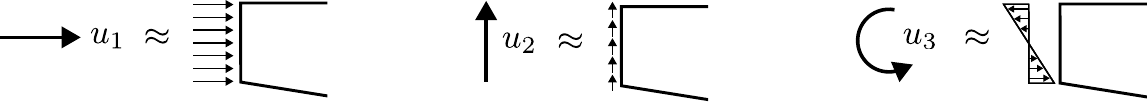}
    \caption{Application of Dirichlet boundary conditions on the local-scale model, shown in 2D. The pointwise axial displacements (left), shear displacements (middle) and the bending rotations (right) are imposed on the local-scale model in a distributed manner.}
	\label{fig:distributed+loads}
\end{figure}

The substructure stiffness matrices $\boldsymbol{K}_{k \times k}$ (\cref{eq:substructure+stiffness}) of the structural nodes are assembled together with the stiffness matrices of the beam elements $\boldsymbol{K_{e}}$ (\cref{eq:timo+beam}) to form the global stiffness matrix $\boldsymbol{K}$, which is then solved for the coefficients $\boldsymbol{\hat u}$ as in~\cref{eq:k+u+f}. Together with~\cref{eq:shape+functions}, this solution represents the displacements and rotations in the global-scale model, which can be mapped back to the local-scale model. To this end, the solution at the boundary degrees of freedom are imposed on the 3D model based on the superposition principle. The zero dimensional displacements and rotations are applied in a distributed manner on the boundary surfaces of the local models, as shown in~\cref{fig:distributed+loads}.

	\section{Numerical Examples} \label{sec:examples}

This section illustrates the proposed multi-scale approach on two examples. \Cref{sec:cantilever} demonstrates and verifies the workflow for a simple cantilever beam under 3D loading. The two-scale model of the beam is verified against a classical Timoshenko beam model in the global scale. In~\Cref{sec:complextree}, the multi-scale approach is applied to analyze a tree canopy structure with multiple geometrically complex structural nodes, which is an ideal candidate for additive manufacturing, yet particularly challenging to analyze and design with conventional tools. 
\subsection{Cantilever Beam} \label{sec:cantilever}
\begin{figure}[h!]
	\hspace{1cm}
	\centering
	\includegraphics[width=0.95\textwidth]{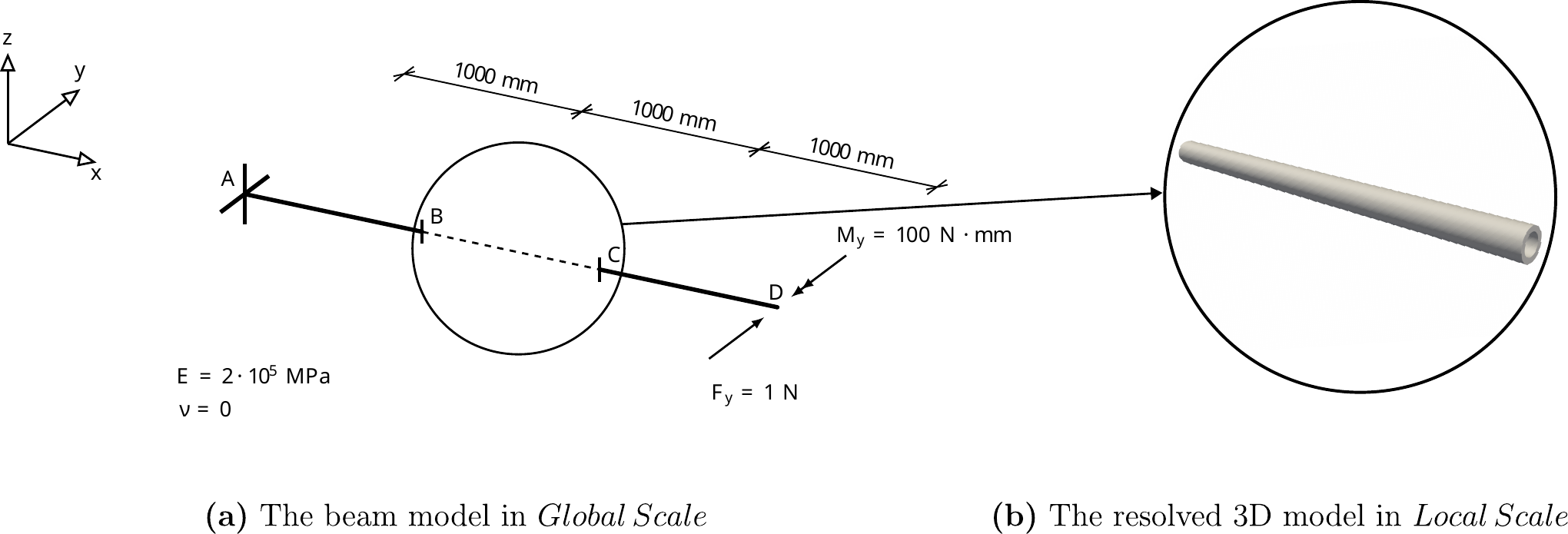}
	\caption{{The problem setup for the cantilever beam.}}
	\label{fig:beam+setup}
\end{figure}

The problem setup for the 3D cantilever beam example is illustrated in~\cref{fig:beam+setup}. The two-scale modeling approach is adopted, where the most simple variant of a structural node with only two connections is chosen as the local-scale (between B and C in~\cref{fig:beam+setup}) and represented by its stiffness values in the global-scale. The beam is clamped at point A and is subjected to a lateral force of 1 \si{N} and a bending moment of 100 \si{N.mm} at point D. A circular beam section with a radius of 30 \si{mm} is assumed for the segment AB and CD, whereas the segment BC has a hollow cross-section with inner and outer radii of 20 and 30 \si{mm}, respectively, as depicted in~\cref{fig:beam+setup}. The global model is discretized by 3D Timoshenko beam elements with a length of 100 \si{mm} per element, while the nodes corresponding to the points B and C are coupled by a substructure following the static condensation procedure outlined in~\Cref{sec:substructure-analysis}. A reference model is set up by modeling the entire structure with Timoshenko beam elements, as in~\Cref{sec:beam-model}, with the same resolution of 100 \si{mm} per element. The reference model properly considers the modified cross-sectional properties of the segment BC. It is widely accepted that the reference model has sufficient approximation quality for slender beams such as the one in question here~\cite{connor2013}. 

\begin{figure}[h!]
  \centering
  \begin{minipage}[c]{0.65\linewidth} 
    \vspace{0pt}
    \includegraphics[width=\linewidth, height=0.53\linewidth]{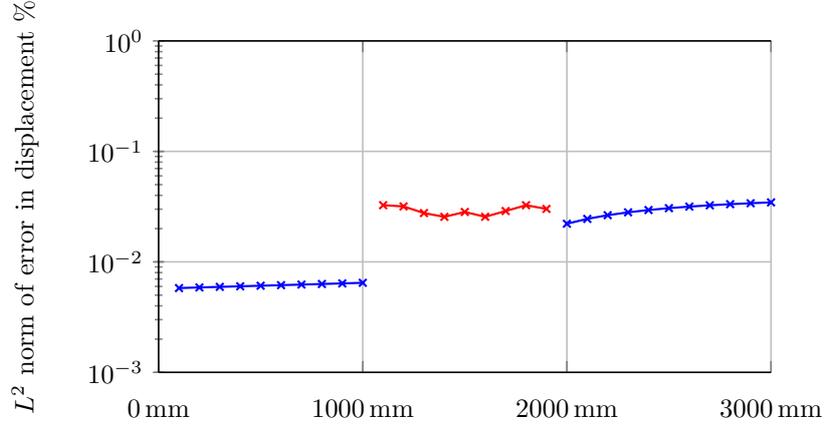}
  \end{minipage}%
    \caption{{$L^{2}$ norm of error in displacement along the beam: the error in local scale shown in red and the error in global scale is shown in blue. }}
\label{fig:beam+l2+error}
\end{figure}

\begin{figure}[h]
	\centering
	\includegraphics[width=1.0\textwidth]{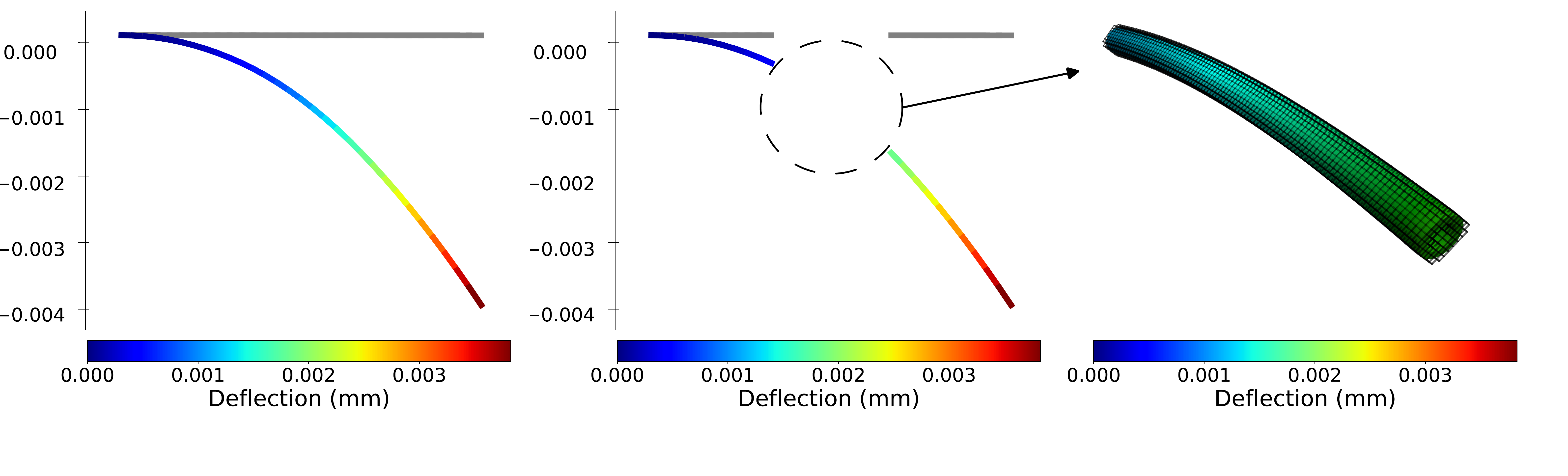}
    \caption{{Deflection of the simple beam in z-direction: the reference solution (left), the global scale solution using the two-scale approach (middle) and the corresponding local scale solution with the FCM mesh (right).}}
	\label{fig:beam+deflections}
\end{figure}

The accuracy of the two-scale approach is measured in the global scale following the error in pointwise $L^2$ norm:

\begin{equation}
    e(x)= \frac{\sqrt{\left(u_{ref}\left(x \right) - u_{ts}\left(x\right) \right)^2}}{\left| u_{ref}\left(x\right)\right|},
\end{equation}
where $u_{ref}\left(x \right)$ is the displacement field computed by the reference model and $u_{ts}\left(x\right)$ is the displacement field computed by the two-scale model. \cref{fig:beam+l2+error} plots the error $e$ along the x-axis of the beam. The error remains below $0.05\%$ throughout the length of the beam. The error is negligible compared to the errors introduced by other assumptions, such as the assumption of clamped ends, which demonstrates that the two-scale approach provides sufficient accuracy for engineering applications. \cref{fig:beam+deflections} depicts the resulting global deflections in z-direction for the reference and the two-scale model. 

\subsection{Complex tree canopy structure} \label{sec:complextree}
\begin{figure}[h]
	\centering
	\includegraphics[width=1\textwidth]{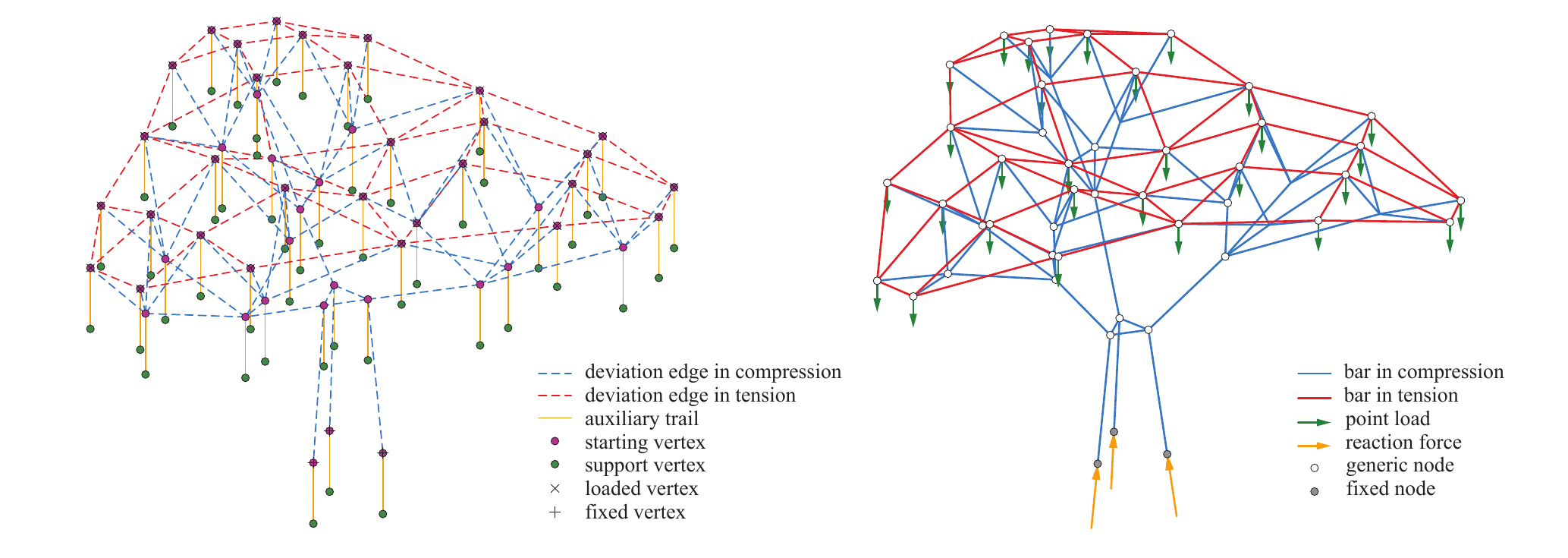}
    \caption{{Combinatorial Equilibrium Modeling (CEM)~\cite{ohlbrock2020,pastrana2022}: topology diagram (left) and structural form in static equilibrium after the form-finding process (right)}}
	\label{fig:cem+diagram}
\end{figure}

The Combinatorial Equilibrium Modeling (CEM)~\cite{ohlbrock2020} method is used to design the initial geometry of the tree canopy structure at the global scale. The CEM is an equilibrium-based form-finding approach grounded in graph theory and Vector-­based Graphic Statics \cite{dacunto2019}. It is particularly tailored for generating spatial pin-jointed frameworks subject to tensile and compressive axial forces according to user-defined initial design inputs: the topology diagram and its associated metric parameters. The topology diagram is a graph representation of the structure (\cref{fig:cem+diagram}). It is composed of vertices, corresponding to the nodes of the structure, and edges, corresponding to the bars of the structure. Vertices are classified into starting, generic, and support vertices, while edges are divided into trail edges and deviation edges. The metric parameters comprise the position of the nodes of the structure corresponding to the starting vertices of the topology diagram, the lengths of the bars corresponding to the trail edges, and the magnitudes of the forces in the bars corresponding to the deviation edges. In the standard formulation of the CEM~\cite{ohlbrock2020}, the form of the structure in static equilibrium is generated sequentially, node-by-node, from the nodes corresponding to the starting vertices up to the nodes corresponding to the support vertices. If specific geometric constraints are imposed onto the structural form (e.g. fixed nodes), an optimization procedure is applied \cite{ohlbrock2017}. In the extended version of the CEM~\cite{pastrana2022}, auxiliary trail edges can be introduced in the topological diagram. These elements correspond to residual forces applied to the structure, which are minimized during the form-finding process (e.g. through gradient-based optimization via automatic differentiation). To generate the initial geometry of the tree canopy at the global scale, point loads of 0.5 \si{kN} are applied to the canopy's top nodes (a total of 15 \si{kN}) while the nodes' positions at the base are fixed (\cref{fig:cem+diagram}). Auxiliary trails are introduced at every vertex of the topology diagram. The form-finding is performed using COMPAS CEM~\cite{pastrana2021a}.

\begin{figure}[h]
	\centering
	\includegraphics[width=1\textwidth]{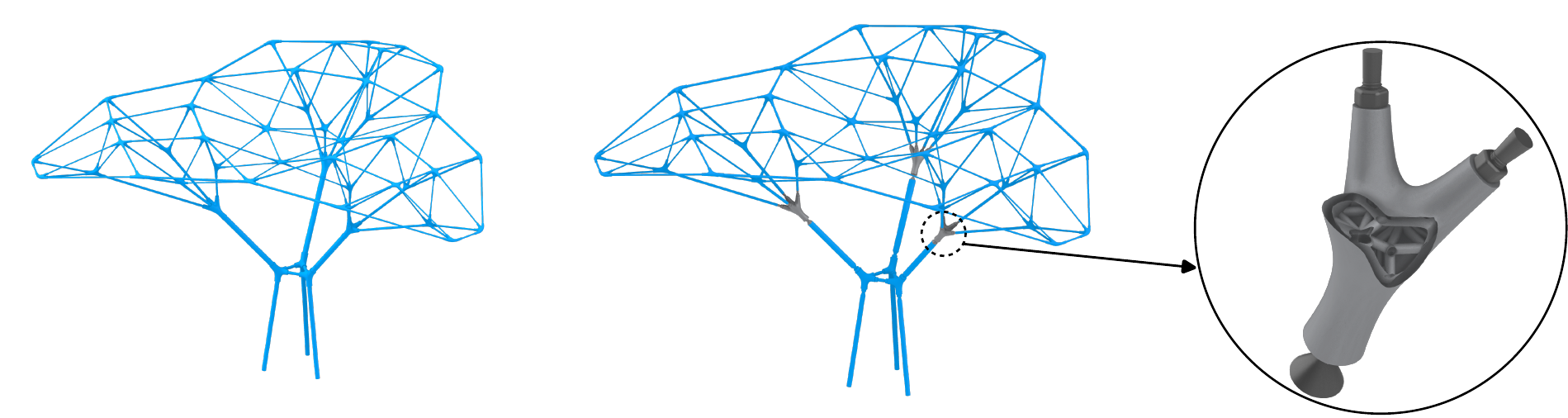}
    \caption{{ The initial design from the CEM form-finding (left) and the refined design with shell-infill composite nodes and bolted connections (middle and right) }}
	\label{fig:initial+design}
\end{figure}

The resulting global design is illustrated in~\cref{fig:initial+design}. The height of the structure is 5 meters and the bars have a circular cross-section with radii ranging from 14 \si{mm} to 50 \si{mm} depending on the resulting internal loads from the form finding. The buckling of the bars in compression is taken into account by making sure that the compressive loads experienced by the bars stay below the corresponding Euler critical buckling load~\cite{goodno2018}.

\begin{figure}
	\centering
	\includegraphics[width=0.45\textwidth]{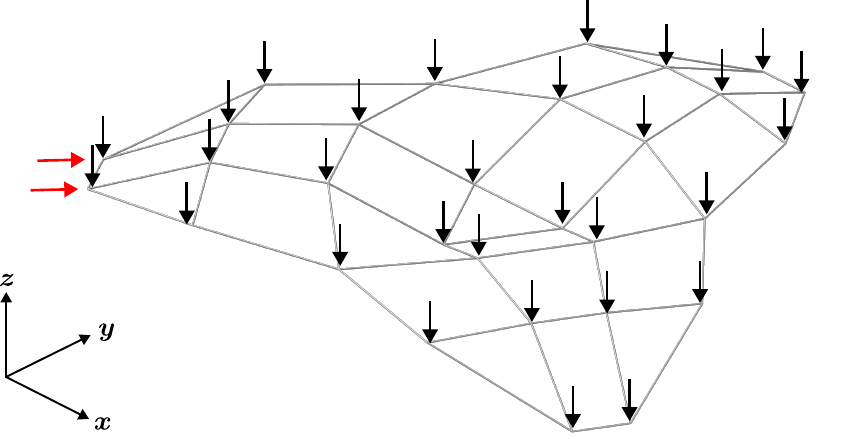}
    \caption{{The global model (shown here in top view) is subjected to a combination of compressive (black) and shear (red) forces. A total of 15 \si{kN} of compressive load and 0.2 \si{kN} of shear load is exerted on the structure.} }
	\label{fig:design+loads}
\end{figure}

The initial design of the structure was based on axial loading of bar elements. However, the behavior of the structure under eccentric design loads, as depicted in~\cref{fig:design+loads}, will be examined in the subsequent analysis using beam elements, considering the relevant geometric scales.

Subsequently, the local-scale structural node models for the selected three nodes are constructed, as represented by the gray color in~\cref{fig:initial+design}. Although the focus of this demonstration is on the selected structural nodes, the proposed workflow is equally applicable to any other complex structural nodes within the structure.

At the early design stage, performing a complete topology optimization for the structural nodes is a challenging task, especially due to the multi-criteria nature of the problem where the mechanical performance, manufacturability, functional integration and other relevant aspects have to be considered. Therefore, we pursue a generative design approach based on shell-infill structures which are commonly used in additive manufacturing \cite{mitchell2018} and are known to facilitate lightweight designs with favorable mechanical properties \cite{duplessis2019, panesar2018, cheng2018}. The design is composed of a solid exterior shell and a open inner structure where parameters including infill density, shell thickness and infill pattern control the design. We employ the Algorithmic Engineering platform Hyperganic$^{\text{TM}}$ Core \cite{2022} to generate feasible designs that incorporate the shell-infill composition. Additionally, the required beam-to-node connections are considered at this scale, where a simple bolted connection is developed, as shown in~\cref{fig:initial+design}, and the size of the connection is considered to be an additional design parameter. The connections are modeled in this scale to better incorporate the influence of the connection, as the rigidity of the connections can strongly influence the global behaviour \cite{hadianfard2003}. It should be noted that the generative modeling workflow is set up in such a way that a node variant, such as the one shown in~\cref{fig:initial+design} that is composed of shell-infill structure with bolt connections, can directly be generated from the initial global-scale design given the geometric input parameters. The generation of a node variant in STL (Standard Triangle Language) format takes approximately 5 minutes to complete. 

\begin{figure}
	\centering
	\includegraphics[width=0.85\textwidth]{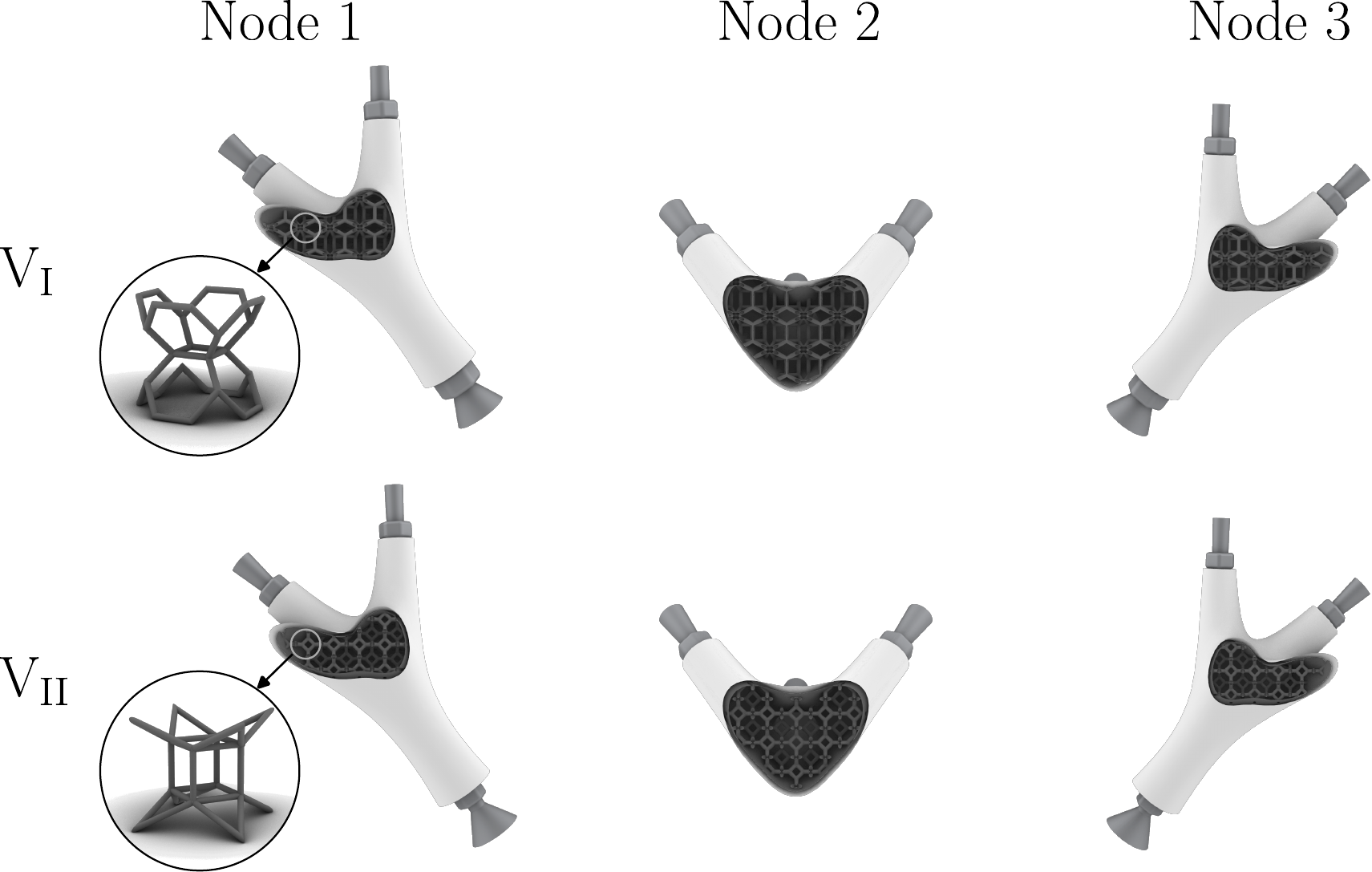}
    \caption{The three structural nodes and the corresponding designs $\text{V}_{\text{I}}$ and $\text{V}_{\text{II}}$.}
	\label{fig:generative+designed+nodes}
\end{figure}

Two sets of design parameters are considered for the local-scale analysis of the selected nodes. The first design alternative, referred to as $\text{V}_{\text{I}}$, specifies a shell thickness of 6 \si{mm} and a volume ratio of approximately 0.41. The second alternative, $\text{V}_{\text{II}}$, sets a shell thickness of 7 \si{mm} and an approximate volume ratio of 0.39. Additionally, the connections of the $\text{V}_{\text{I}}$ variants are designed to be larger than the $\text{V}_{\text{II}}$ variants. Consequently, the node variants $\text{V}_{\text{I}}$ have approximately $5\%$ more mass than the corresponding node variants $\text{V}_{\text{II}}$. The design alternatives with the corresponding infill patterns and connections are depicted in~\cref{fig:generative+designed+nodes}, where the upper and lower row illustrates the node designs following the designs $\text{V}_{\text{I}}$ and $\text{V}_{\text{II}}$, respectively. In short, each of the selected three nodes has a $\text{V}_{\text{I}}$ and a $\text{V}_{\text{II}}$ variant, resulting in a total of six distinct node models.

The first step of the two-scale analysis is the computation of the substructure stiffness matrices $\boldsymbol{K}_{k \times k}$ of the corresponding structural nodes. To this end, the full dimensional unconstrained stiffness matrix $\boldsymbol{K}_{n \times n}$ is computed using the FCM for each node variant. Subsequently, we perform $k$ separate simulations per node variant to compute the corresponding change of basis matrices $\boldsymbol{N}_{n \times k}$. The number of interfacing beam degrees of freedom is $k=30$ as the nodes are connected to five beam elements with six boundary DOF each. The substructure stiffness matrices are then calculated following~\cref{eq:substructure+stiffness}. To correctly resolve the fine geometric details of the nodes, FCM discretizations of $110 \times 90 \times 132$ (for Node 1 and 3) and $91 \times 125 \times 133$ (for Node 2) are used and the integrated Legendre polynomials spanning each element have a polynomial degree of 3, resulting in $\backsim 2.10^{6}$ degrees of freedom. Moreover, an octree integration subdivision depth of 4 is chosen, the indicator function $\alpha$ is set to $10^{-10}$ and the penalty parameter $\beta$ is set to $10^{14}$.
  
The substructure analysis is conducted on a dual socket computer with Intel$^{\text{TM}}$ Xeon$^{\text{TM}}$ Gold 6230 processors, with a total of 80 logical cores running at a clock speed of 2.1 GHz. The computation of the substructure stiffness matrix of a structural node, based on the simulation parameters listed above, has an execution time of $\backsim\text{2.5 hours}$. This duration includes solving the linear system $k=30$ times and performing the change of basis operations. A significant portion of the computational time is spent on numerical integration over the complex domain, and the use of advanced techniques can greatly improve the efficiency of this process, e.g. \cite{Kudela2016, garhuom2022}. Note that the process of generating a node design and conducting a subsequent substructure analysis is automated, eliminating the need for manual engineering effort to perform the analysis on the complex node model.

We introduce the global-scale models that incorporate the two node designs $\text{V}_{\text{I}}$ and $\text{V}_{\text{II}}$. The models are set up with 985 Timoshenko beam elements and a Young's modulus of $E=2.0\cdot 10^5$ $\text{MPa}$. In the first case, the substructure matrices corresponding to the node variants $\text{V}_{\text{I}}$ are assembled together with the beam elements. The resulting model is analyzed under the specified design loads (\cref{fig:design+loads}). To demonstrate the versatility of the approach, a second case is analyzed, in which the substructure stiffnesses of node variants $\text{V}_{\text{II}}$ are assembled with the beam elements, while all other modeling and analysis parameters remain unchanged. The resulting displacements and moments of the corresponding global systems with node designs $\text{V}_{\text{I}}$ and $\text{V}_{\text{II}}$ are illustrated in~\cref{fig:global+deflections} and~\cref{fig:global+moments}. It can be seen that the proposed approach facilitates the evaluation of the large-scale mechanical response, taking into account the impact of local design decisions.

\subsubsection{The global-scale response}  \label{sec:global+results}

\begin{figure}
	\centering
	\includegraphics[width=0.85\textwidth]{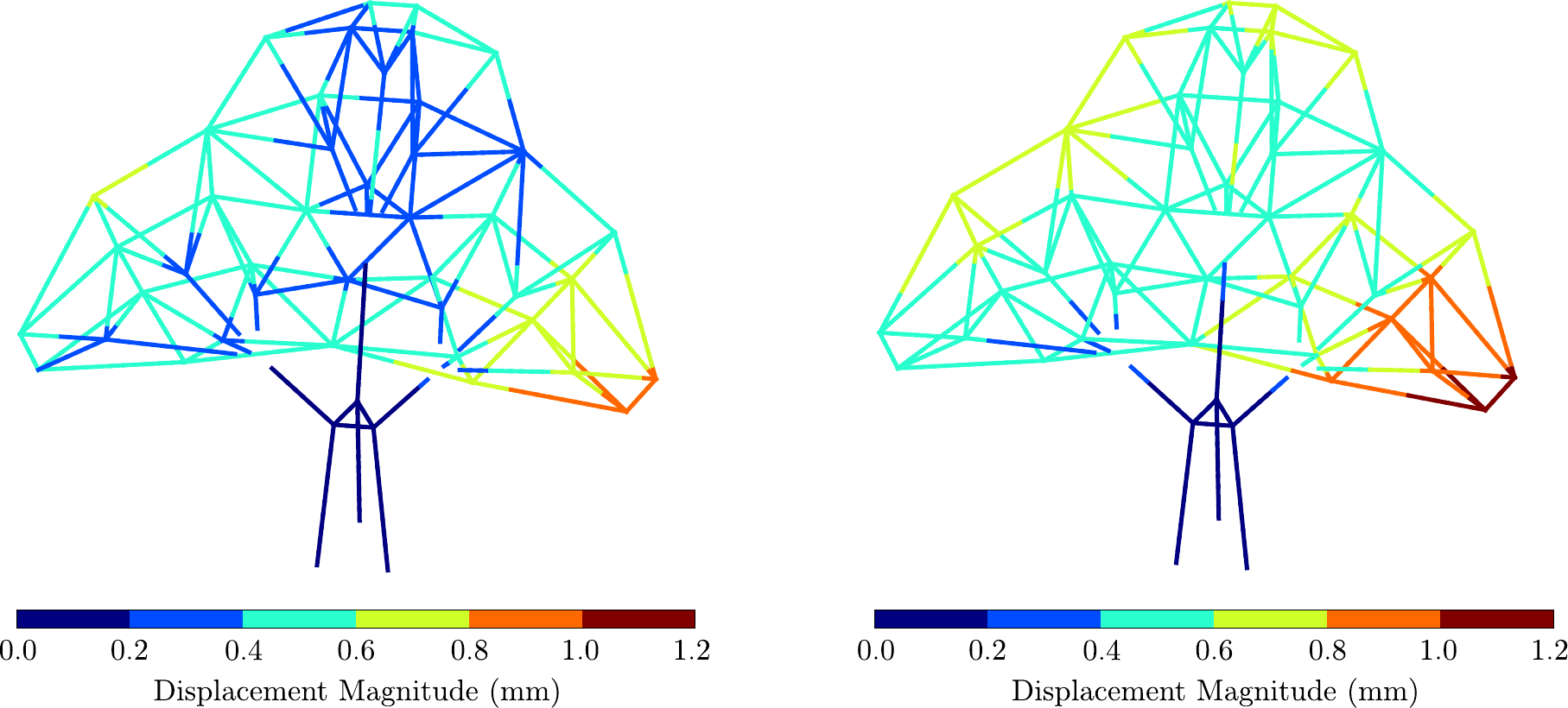}
    \caption{The displacements from the global analysis: the deformation of the global system with nodes $\text{V}_{\text{I}}$ (left) and with nodes $\text{V}_{\text{II}}$ (right).}
	\label{fig:global+deflections}
\end{figure}

\begin{figure}
	\centering
	\includegraphics[width=0.85\textwidth]{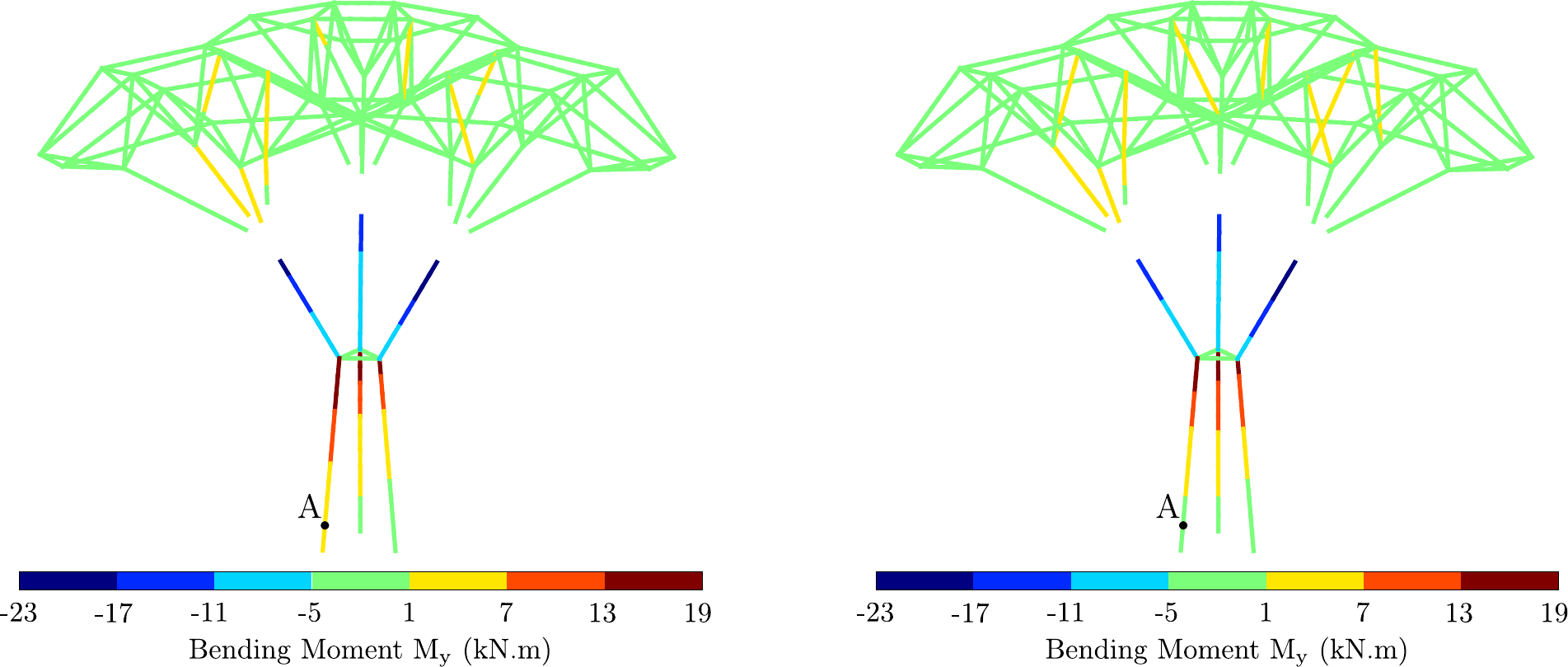}
    \caption{The bending moments from the global analysis: the bending moment $\text{M}_\text{y}$ of the global system with nodes $\text{V}_{\text{I}}$ (left) and with nodes $\text{V}_{\text{II}}$ (right).}
	\label{fig:global+moments}
\end{figure}

The initial global analysis considering the first set of node designs $\text{V}_{\text{I}}$ indicates a maximum nodal displacement of 0.9 \si{mm}, whereas the maximum displacement increases to more than 1.1 \si{mm} when the global model incorporates the node variants $\text{V}_{\text{II}}$, as depicted in~\cref{fig:global+deflections}. In fact, the second node variants cause a greater overall deformation of the structure. This considerable increase in displacements (more than 15\%) can be attributed to the lower connection stiffness and different shell-infill composition. On the other hand, the different node designs do not significantly impact the internal bending moment $\text{M}_{\text{y}}$, as shown in~\cref{fig:global+moments}. This can be attributed to the redistribution of the internal forces which is caused by different stiffnesses of the two node designs. For example, the point A in~\cref{fig:global+moments} (left) experiences an internal bending moment $\text{M}_{\text{y}}$ of approximately 3 \si{kN.m} when the node designs $\text{V}_{\text{I}}$ are used in the global model, whereas $\text{M}_{\text{y}}$ at point A takes a value of approximately -1 \si{kN.m} if the node variants $\text{V}_{\text{II}}$ are incorporated, as shown in~\cref{fig:global+moments} (right). This behaviour is a common response that can be attributed to the statical indeterminacy of the structure \cite{connor2013}.

\begin{figure}
	\centering
	\includegraphics[width=0.95\textwidth]{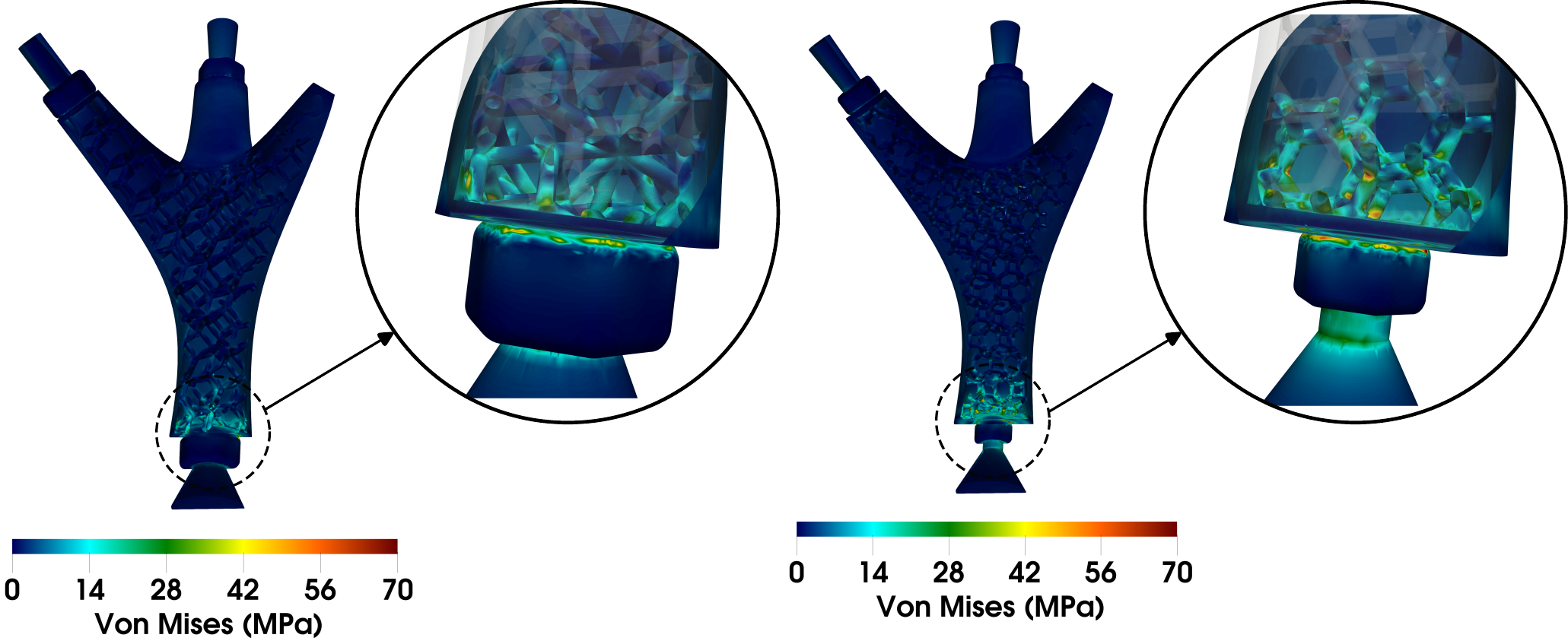}
    \caption{The results of the local stress analysis illustrating the stress state at the most critical end of the connections for the $\text{V}_{\text{I}}$ (left) and $\text{V}_{\text{II}}$ (right) variants of Node 3.}
	\label{fig:local+stress+1}
\end{figure}

\subsubsection{The local-scale response}  \label{sec:local+results}
The global solution at the boundaries of the substructure is then mapped back to the local-scale models, following~\cref{sec:global-to-local}, to analyze the resolved stress state at the local level. Based on the results of the global-scale analysis, it can be observed that the $\text{V}_{\text{II}}$ variant of the nodes experience significantly more deformations. \cref{fig:local+stress+1} illustrates the results of the local-scale stress analysis on the two Node 3 designs (\cref{fig:generative+designed+nodes}), where the stresses are post-processed on the corresponding STL surfaces. It indicates that the higher deformations experienced by the $\text{V}_{\text{II}}$ variant result in higher local stresses. More specifically, the maximum stress experienced by the $\text{V}_{\text{I}}$ variant stays below 60 \si{MPa}, while the $\text{V}_{\text{II}}$ variant is subjected to stresses exceeding 70 \si{MPa}.

\begin{remark}
\cref{fig:local+stress+1} illustrates the interior section of the two structural node variants and the geometries are represented using their boundary surfaces (STL). Note that the section views do not fully reveal the solid infill within the nodes, as the infill is only defined by its boundary. This is purely a post-processing artifact.
\end{remark}

	\section{Conclusion} \label{sec:conclusion}
The present work proposes a two-scale approach for the design and analysis of space frame structures with additively manufactured parts, where the emphasis is put on strong automation capabilities. In this approach, the global scale model assumes 1D linear beam elements, whereas AM structural nodes are modeled as substructures based on fully resolved 3D models. The relevant mechanical characteristics of the structural nodes are numerically reduced using an automated dimensional reduction process. The reduced stiffness quantities are assembled in the global model, allowing for efficient structural analysis at the global scale. The global solution is imposed on the local-scale models, enabling local stress state analysis of the structural nodes using the finite cell method (FCM). This multi-scale engineering design and analysis workflow benefits from the use of the FCM, as it reduces the amount of manual engineering labor required to analyze the complex AM components in the local-scale.

The proposed approach is first verified on a cantilever beam example where it is shown that the methodology introduces a displacement error of $0.05\%$, which provides sufficient accuracy for engineering applications. The advantages of the approach are demonstrated using a tree canopy structure with AM structural nodes. The global design of the structure is provided using a form-finding approach based on the Combinatorial Equilibrium Modeling framework, while the local structural nodes are generatively designed to explore alternative feasible designs. Finally, it is shown that the methodology can be used to analyze the mechanical response of the structure such that both local and global behavior can be assessed depending on the design decisions.  

The current work is restricted to linear-elastic analysis of the space-frames. Further research is needed to extend the methodology to cover the non-linear response of such structures, taking into account the relevant scales. This is particularly important as ductile materials such as steel are commonly used in space-frame construction. Moreover, the computational time associated with the substructure analysis can be reduced by utilizing advanced numerical integration techniques \cite{Kudela2016,garhuom2022}.

Furthermore, the geometrical defects that can be induced by the additive manufacturing process are not yet considered in the local-scale. This difference in geometry between the as-designed and as-manufactured products has been shown to affect the mechanical behaviour of AM parts \cite{korshunova2021,korshunova2021a}. Thus, it is an interesting research topic to incorporate the precise as-manufactured geometry of structural nodes in the analysis workflow and investigate their mechanical effects at relevant scales.

\section*{Acknowledgements} 
We gratefully acknowledge the support of Deutsche Forschungsgemeinschaft (DFG, Germany) through the project 414265976 TRR 277 C-01. We also thank Tao Sun for contributing to the development of the tree canopy model and Hyperganic for providing us with an educational license to their Hyperganic Core platform.

\bibliographystyle{ieeetr}
\bibliography{library}

\end{document}